\documentclass[12pt]{article}
\usepackage{amsmath}
\usepackage{graphicx}
\usepackage{float}
\usepackage{caption2}

\begin{document}

\author{Jos\'{e} L. Cereceda \\
\textit{C/Alto del Le\'{o}n 8, 4A, 28038 Madrid, Spain}}
\title{\textbf{Quantum perfect correlations and  \\
\vspace{-2mm} Hardy's nonlocality theorem}\thanks{This paper has been originally
published in: J.L. Cereceda, \textit{Found. Phys. Lett.} \textbf{12}(3),
211-231 (1999).}}
\date{August 12, 1999}
\maketitle

\begin{abstract}
In this paper the failure of Hardy's nonlocality proof for the class of
maximally entangled states is considered. A detailed analysis shows that the
incompatibility of the Hardy equations for this class of states physically
originates from the fact that the existence of quantum perfect correlations
for the three pairs of two-valued observables $(D_{11},D_{21})$, $(D_{11},
D_{22})$, and $(D_{12},D_{21})$ [in the sense of having with certainty equal
(different) readings for a joint measurement of any one of the pairs $%
(D_{11},D_{21})$, $(D_{11},D_{22})$, and $(D_{12},D_{21})$], necessarily
entails perfect correlation for the pair of observables $(D_{12},D_{22})$
[in the sense of having with certainty equal (different) readings for a
joint measurement of the pair $(D_{12},D_{22})$]. Indeed, the set of these
four perfect correlations is found to satisfy the CHSH inequality, and then
no violations of local realism will arise for the maximally entangled state
as far as the four observables $D_{ij}$, $i,j=1$ or $2$, are concerned. The
connection between this fact and the impossibility for the quantum
mechanical predictions to give the maximum possible theoretical violation of
the CHSH inequality is pointed out. Moreover, it is generally proved that
the fulfillment of all the Hardy nonlocality conditions necessarily entails
a violation of the resulting CHSH inequality. The largest violation of this
latter inequality is determined.

\vspace{.3cm}
\noindent \samepage{\textit{Key words:} perfect correlation, maximally
entangled state, local realism, Hardy's nonlocality theorem, Bell's
inequality.}
\end{abstract}

\section{Introduction}

A very remarkable feature of Hardy's nonlocality proof [1] is that it goes
through for any entangled states of a $2\times 2$ system except those which
are maximally entangled such as the singlet state of two spin-$\frac{1}{2}$
particles. At first sight this might appear rather surprising in view of the
fact that maximally entangled states yield the maximum violation predicted
by quantum mechanics of the Clauser-Horne-Shimony-Holt (CHSH) inequality
[2]. Referring himself to this failure, Hardy states that [1], ``the reason
for this is that the proof relies on a certain lack of symmetry that is not
available in the case of a maximally entangled state.'' Indeed, it has been
found [1,\,3-6] that the set of Hardy equations (see Eqs.\ (8a)-(8d) below)
upon which the nonlocality contradiction is constructed is incompatible for
the case of maximal entanglement in the sense that for this case the
fulfillment of all three conditions (8a), (8b), and (8c) precludes the
fulfillment of condition (8d), and vice versa. From a mathematical point of
view this is the reason for the failure, and this would be the end of the
story.

In this paper I would like to account for this failure from a somewhat
different perspective which provides a fuller mathematical understanding of
the structure of Hardy's theorem. This will allow us to gain some insight
into the physical cause of the inability of completely entangled states to
produce a Hardy-type nonlocality contradiction. So, after introducing in
Sec.\ 2 some general results concerning the conditions needed to achieve
perfect correlations for $2\times 2$ systems, in Sec.\ 3 it will be shown
that the fulfillment of all three conditions (8a)-(8c) in the case of a
maximally entangled state necessarily entails perfect correlation between
the two measurement outcomes (one for each particle) obtained in \textit{any}
one of the four possible combinations of joint measurements ($D_{1k},D_{2l}$),
$k,l=1$ or $2$, one might actually perform on both particles, where $D_{1k}$
and $D_{2l}$ are single-particle observables associated with particles 1 and 2,
respectively. However, as we shall see, the CHSH inequality is
fulfilled for such maximal-entanglement-induced perfect correlations, and,
thereby, no violations of local realism will arise for the maximally
entangled state as long as the four observables $D_{1k}$ and $D_{2l}$
involved in the CHSH inequality make conditions (8a), (8b), and (8c) hold.
Indeed, the fulfillment of the CHSH inequality for such perfect correlations
means that all of them can be consistently explained in terms of a local
hidden-variable model (see, for instance, Appendix D in Ref.\ 7 for an
explicit example of such a model that accounts for the perfect correlations
of two spin-$\frac{1}{2}$ particles in the singlet state). This is
ultimately the physical reason why Hardy's nonlocality argument does not
work for the maximally entangled case. Moreover, as will become clear, the
failure of Hardy's argument for the maximally entangled state is,
interestingly enough, closely related to the fact that the quantum
mechanical predictions cannot give the maximal possible theoretical
violation of the CHSH inequality. In Sec.\ 4, the general case of
less-than-maximally entangled state is considered, and it is shown how the
fulfillment of all the Hardy conditions (8a)-(8d) necessarily leads to a
violation of the resulting CHSH inequality. The largest extent of this
violation is determined. Finally, in Sec.\ 5, examples are given
illustrating the fact that maximally entangled states yield the maximum
quantum mechanical violation of the CHSH inequality, while this inequality
is necessarily obeyed for such states if these latter are constrained to
satisfy the conditions (8a), (8b), and (8c).

\section{Perfect correlations for $\mathbf{2\times 2}$ systems}

Hardy's nonlocality proof involves an experimental set-up of the
Einstein-Podolsky-Rosen-Bohm type [8,\,9]: two correlated particles 1 and 2
fly apart in opposite directions from a common source such that each of them
subsequently impinges on an appropriate measuring device which can measure
either one of two physical observables at a time---$D_{11}$ or $D_{12}$ for
(the apparatus measuring) particle 1, and $D_{21}$ or $D_{22}$ for (the one
measuring) particle 2. Conventionally, it is supposed that the measurement
of each one of these observables gives the possible outcomes ``$+1$'' and
``$-1$'' (this is the case that arises, for example, in the realistic situation
[10] in which a photon is detected behind a two-channel polarizer, with the
value $+1$ $(-1)$ assigned to detections corresponding to the transmitted
(reflected) photon\footnote{%
Of course in a real experiment it may well happen that neither one of the
two photons of a given pair emitted by the source is registered by the
detection system (or else that only one of them is detected), even though
the two photons have nearly opposite directions. This is mainly
due (although not exclusively) to the low efficiency of the available
detectors. The usual way of circumventing this problem is to assume that the
subensemble of actually detected pairs is a representative sample of the whole
ensemble of emitted photon pairs.}), so that we shall generally assume that
the operators associated with such observables are of the form, $\hat{D}_{ij}=
\left| d_{ij}^{+}\right\rangle \left\langle d_{ij}^{+}\right| -\left|
d_{ij}^{-}\right\rangle \left\langle d_{ij}^{-}\right| $, with $i,j=1$ or $2$%
, and where $\left\{ \left| d_{ij}^{+}\right\rangle ,\left|
d_{ij}^{-}\right\rangle \right\} $ constitutes an orthonormal basis for the
Hilbert space pertaining to particle $i$. On the other hand, according to
the Schmidt decomposition theorem (see, for instance, Ref.\ 11 for a
recent account of this topic), we can always write the quantum pure
state of our two-particle system as a sum of two biorthogonal terms 
\begin{equation}
\left| \eta \right\rangle =c_{1}\left| u_{1}\right\rangle \left|
u_{2}\right\rangle +c_{2}\left| v_{1}\right\rangle \left| v_{2}\right\rangle ,
\end{equation}
for some suitably chosen orthonormal basis $\{\left| u_{i}\right\rangle
,\left| v_{i}\right\rangle \}$ for particle $i$, with the real coefficients $%
c_{1}$ and $c_{2}$ satisfying the relation $c_{1}^{2}+c_{2}^{2}=1$. Now, by
expressing the eigenvectors $\left| d_{ij}^{+}\right\rangle $ and $\left|
d_{ij}^{-}\right\rangle $ in terms of the basis vectors $\left|
u_{i}\right\rangle $ and $\left| v_{i}\right\rangle $%
\begin{align}
\left| d_{ij}^{+}\right\rangle &=e^{i\alpha _{ij}}\cos \beta _{ij}\left|
u_{i}\right\rangle +e^{i\gamma _{ij}}\sin \beta _{ij}\left|
v_{i}\right\rangle ,  \tag{2a} \\
\left| d_{ij}^{-}\right\rangle &=-e^{-i\gamma _{ij}}\sin \beta _{ij}\left|
u_{i}\right\rangle +e^{-i\alpha _{ij}}\cos \beta _{ij}\left|
v_{i}\right\rangle ,  \tag{2b}
\setcounter{equation}{2}
\end{align}
one can evaluate the quantum probability distributions $P_{\eta
}(D_{1k}=m,D_{2l}=n)$, with $m,n=\pm 1$ and $k,l=1$ or $2$, that a joint
measurement of the observables $D_{1k}$ and $D_{2l}$ on particles 1 and 2,
respectively, gives the outcomes $D_{1k}=m$ and $D_{2l}=n$ when the
particles are described by the state vector (1). These are given by
\pagebreak
\begin{align}
P_{\eta }(D_{1k} &=+1,D_{2l}=+1)  \nonumber \\
&=c_{1}^{2}\cos ^{2}\beta _{1k}\cos ^{2}\beta _{2l}+c_{2}^{2}\sin ^{2}\beta
_{1k}\sin ^{2}\beta _{2l}  \nonumber \\
&\;\;\;\;+\tfrac{1}{2}c_{1}c_{2}\cos \delta _{1k2l}\sin 2\beta _{1k}\sin 2
\beta_{2l}\, ,  \tag{3a}
\end{align}
\begin{align}
P_{\eta }(D_{1k} &=-1,D_{2l}=-1)  \nonumber \\
&=c_{1}^{2}\sin ^{2}\beta _{1k}\sin ^{2}\beta _{2l}+c_{2}^{2}\cos ^{2}\beta
_{1k}\cos ^{2}\beta _{2l}  \nonumber \\
&\;\;\;\;+\tfrac{1}{2}c_{1}c_{2}\cos \delta _{1k2l}\sin 2\beta _{1k}\sin 2
\beta_{2l}\, ,  \tag{3b}
\end{align}
\begin{align}
P_{\eta }(D_{1k} &=+1,D_{2l}=-1)  \nonumber \\
&=c_{1}^{2}\cos ^{2}\beta _{1k}\sin ^{2}\beta _{2l}+c_{2}^{2}\sin ^{2}\beta
_{1k}\cos ^{2}\beta _{2l}  \nonumber \\
&\;\;\;\;-\tfrac{1}{2}c_{1}c_{2}\cos \delta _{1k2l}\sin 2\beta _{1k}\sin 2
\beta_{2l}\, ,  \tag{3c}
\end{align}
\begin{align}
P_{\eta }(D_{1k} &=-1,D_{2l}=+1)  \nonumber \\
&=c_{1}^{2}\sin ^{2}\beta _{1k}\cos ^{2}\beta _{2l}+c_{2}^{2}\cos ^{2}\beta
_{1k}\sin ^{2}\beta _{2l}  \nonumber \\
&\;\;\;\;-\tfrac{1}{2}c_{1}c_{2}\cos \delta _{1k2l}\sin 2\beta _{1k}\sin 2
\beta_{2l}\, ,  \tag{3d}
\setcounter{equation}{3}
\end{align}
where $\delta _{1k2l}=\delta _{1k}-\delta _{2l}$, with $\delta _{1k}=\gamma
_{1k}-\alpha _{1k}$ and $\delta _{2l}=\alpha _{2l}-\gamma _{2l}$. Of course,
the above probabilities add up to unity 
\begin{equation}
\sum_{m,n=\pm 1}P_{\eta }(D_{1k}=m,D_{2l}=n)=1\, .  
\end{equation}
We will note at this point that for the special case in which $\left|
c_{1}\right| =\left| c_{2}\right| =2^{-1/2}$ (i.e., when state (1) happens
to be totally entangled), the following two equalities, $P_{\eta
}(D_{1k}=+1,D_{2l}=+1)=P_{\eta }(D_{1k}=-1,D_{2l}=-1)$ and $P_{\eta
}(D_{1k}=+1,D_{2l}=-1)=P_{\eta }(D_{1k}=-1,D_{2l}=+1)$, hold true. Now, the
expectation value of the product of the measurement outcomes of $D_{1k}$ and 
$D_{2l}$ is defined (in an obvious notation) as 
\begin{equation}
E_{\eta }(D_{1k},D_{2l})=P_{\eta }^{++}+P_{\eta }^{--}-P_{\eta
}^{+-}-P_{\eta }^{-+}.  
\end{equation}
Substituting expressions (3a)-(3d) into Eq.\ (5) gives 
\begin{equation}
E_{\eta }(D_{1k},D_{2l})=\cos 2\beta _{1k}\cos 2\beta _{2l}+2c_{1}c_{2}\cos
\delta _{1k2l}\sin 2\beta _{1k}\sin 2\beta _{2l}\, .  
\end{equation}
(Note, incidentally, that for either $c_{1}=0$ or $c_{2}=0$ (i.e., for
product states) the quantum correlation function (6) factorizes with respect
to the parameters $\beta _{1k}$ and $\beta _{2l}$, and consequently it will
be unable to yield a violation of Bell's inequality.) In general the
correlation function (6) takes on values in the range between $-1$ and $+1$.
We are interested in determining the conditions under which this function
attains its extremal values $\pm 1$. By direct inspection of Eq.\ (6) it
follows at once that whenever we have $\beta _{1k}=n_{1k}\pi /2$ and $\beta
_{2l}=n_{2l}\pi /2$ ($n_{1k},n_{2l}=0,\pm 1,\pm 2,\ldots\,\, $), perfect
correlations happen for any state of the form (1) irrespective of the values
of $c_{1}$ and $c_{2}$. This corresponds to the case that the operators $%
\hat{D}_{1k}$ and $\hat{D}_{2l}$ have eigenvectors that arise from the
Schmidt decomposition of the entangled state (see Eq.\ (1)). For this case
the operators $\hat{D}_{1k}$ and $\hat{D}_{2l}$ take the form $\hat{D}%
_{1k}=\mu _{1k}\left| u_{1}\right\rangle \left\langle u_{1}\right| +\eta
_{1k}\left| v_{1}\right\rangle \left\langle v_{1}\right| $ and $\hat{D}%
_{2l}=\mu _{2l}\left| u_{2}\right\rangle \left\langle u_{2}\right| +\eta
_{2l}\left| v_{2}\right\rangle \left\langle v_{2}\right| $ (with $\mu _{1k}$%
, $\eta _{1k}$, $\mu _{2l}$, and $\eta _{2l}$ being sign factors fulfilling $%
\mu _{1k}\eta _{1k}=\mu _{2l}\eta _{2l}=-1$), and then, from the very
structure of the state vector (1), it is apparent that a measurement of $%
D_{1k}$ on particle 1 will uniquely determine the outcome of a measurement
of $D_{2l}$ on particle 2, and vice versa. On the other hand, for the case
in which $c_{1},c_{2}\neq 0$, and $\beta _{1k}\neq n_{1k}\pi /2$, $\beta
_{2l}\neq n_{2l}\pi /2$, it follows that in order for the correlation
function (6) to take on its extremal values $\pm 1$ it is necessary that,
(i) the state (1) be maximally entangled, and (ii) $\delta
_{1k2l}=n_{1k2l}\pi $, $n_{1k2l}=0,\pm 1,\pm 2,\ldots $ . Indeed, whenever
the equality $2c_{1}c_{2}\cos \delta _{1k2l}=\mp 1$ holds, Eq.\ (6) reduces
to 
\begin{equation}
E_{\eta }(D_{1k},D_{2l})=\cos 2(\beta _{1k}\pm \beta _{2l})\, ,
\end{equation}
which attains the value $+1$ or $-1$ for $\beta _{1k}\pm \beta
_{2l}=m_{1k2l}\pi /2$, $m_{1k2l}=0,\pm 1$, $\pm 2,\ldots $ .

\section{Hardy's nonlocality conditions: implications
for the maximally entangled case}

In terms of joint probabilities, a generic two-particle state $\left| \eta
\right\rangle $ of the form (1) will show Hardy-type nonlocality
contradiction if the following four conditions are simultaneously fulfilled
for the state $\left| \eta \right\rangle $ [1,\,3-6],\footnote{%
The same type of contradiction is obtained if, in Eqs.\ (8a)-(8d), we convert
all the $+1$'s into $-1$'s, and vice versa. Likewise, the same holds true if
we reverse the sign \textit{only} to the outcomes of $D_{1k}$ for each of
the Eqs.\ (8a)-(8d). Indeed, as may be easily checked, the following set of
conditions 
\begin{align*}
P_{\eta }(D_{11}=+1,D_{21}=-1) &\,=\,0 \, , \\
P_{\eta }(D_{11}=-1,D_{22}=+1) &\,=\,0 \, , \\
P_{\eta }(D_{12}=-1,D_{21}=+1) &\,=\,0 \, , \\
P_{\eta }(D_{12}=-1,D_{22}=+1) &\,>\,0 \, ,
\end{align*}
will also lead to a Hardy-type nonlocality contradiction. Naturally, by
symmetry, the same is true if we reverse the sign \textit{only} to the
outcomes of $D_{2l}$ for each of the Eqs.\ (8a)-(8d).} 
\begin{align}
P_{\eta }(D_{11}=-1,D_{21}=-1) &\,=\,0 \, ,  \tag{8a} \\
P_{\eta }(D_{11}=+1,D_{22}=+1) &\,=\,0 \, ,  \tag{8b} \\
P_{\eta }(D_{12}=+1,D_{21}=+1) &\,=\,0 \, ,  \tag{8c} \\
P_{\eta }(D_{12}=+1,D_{22}=+1) &\,>\,0 \, .  \tag{8d}
\end{align}
Taking into account the aforementioned equalities, $P_{\eta
}(D_{1k}=+1,D_{2l}=+1)=P_{\eta }(D_{1k}=-1,D_{2l}=-1)$ and $P_{\eta
}(D_{1k}=+1,D_{2l}=-1)=P_{\eta }(D_{1k}=-1,D_{2l}=+1)$, which are valid only
in the case that the state $\left| \eta \right\rangle $ is maximally
entangled, it is immediate to see that the fulfillment of Eqs.\ (8a), (8b),
and (8c) for such a special case implies perfect correlation between the
measurement outcomes of $D_{11}$ and $D_{21}$, $D_{11}$ and $D_{22}$, and $%
D_{12}$ and $D_{21}$, respectively. So, for example, if condition (8a) is
fulfilled for the maximally entangled state we have that $P_{\eta
}(D_{11}=-1,D_{21}=-1)=P_{\eta }(D_{11}=+1,D_{21}=+1)=0$. Therefore, the
probability $P_{\eta }^{\neq }(D_{11},D_{21})=P_{\eta
}(D_{11}=+1,D_{21}=-1)+P_{\eta }(D_{11}=-1,D_{21}=+1)$ of getting different
readings for a measurement of $D_{11}$ and $D_{21}$ is unity, and thus the
measurement outcomes for such observables are perfectly correlated in that
whenever the outcome $+1$ $(-1)$ is observed for particle 1 then with
certainty the outcome $-1$ $(+1)$ will be observed for particle 2, and vice
versa. Mathematically, this is expressed by the fact that $E_{\eta
}(D_{11},D_{21})=-1$ (see Eq.\ (5)). A similar conclusion applies to the
measurement outcomes of $D_{11}$ and $D_{22}$, and also to the measurement
outcomes of $D_{12}$ and $D_{21}$, if conditions (8b) and (8c) are to be
satisfied for the maximally entangled state; namely, $E_{\eta
}(D_{11},D_{22})=E_{\eta }(D_{12},D_{21})=-1$.

Now we are going to show that the fulfillment of all three conditions
(8a)-(8c) for the maximally entangled state also implies perfect correlation
for the observables $D_{12}$ and $D_{22}$, in the sense of having in all
cases different measurement outcomes for $D_{12}$ and $D_{22}$. Clearly,
this makes it impossible the fulfillment of the remaining condition in Eq.\
(8d). First of all we note that, as a general rule, the fulfillment of Eqs.\
(8a), (8b), and (8c), requires, respectively, that $\delta
_{1121}=n_{1121}\pi $, $\delta _{1122}=n_{1122}\pi $, and $\delta
_{1221}=n_{1221}\pi $. This is so because the derivative of $P_{\eta
}(D_{1k}=m,D_{2l}=n)$ with respect to the variable $\delta _{1k2l}$ must be
zero at the minimum value $P_{\eta }=0$. For concreteness, and without any
loss of generality, from now on we shall take the choice $%
n_{1121}=n_{1122}=n_{1221}=0$, so that $\delta _{1121}=\delta _{1122}=\delta
_{1221}=0$. Recalling that $\delta _{1k2l}=\delta _{1k}-\delta _{2l}$, this
in turn implies that the relative phases $\delta _{ij}$ are constrained to
obey the relation $\delta _{11}=\delta _{12}=\delta _{21}=\delta _{22}$.
Therefore, we deduce that the angle $\delta _{1222}=\delta _{12}-\delta
_{22} $ must equally be zero. Of course the fact that the cosine function $%
\cos \delta _{1222}$ takes the value $+1$ [or else $-1$] is a necessary
(although not a sufficient) condition in order for the probability in Eq.\
(8d) to reach its minimum value 0. The constraint $\delta _{1222}=0$ having
been established, all what we need to reach the desired conclusion is the
following straightforward mathematical result [6], whose proof is given in
the Appendix.

\vspace{.5cm}
\noindent
\rule{\textwidth}{.15mm}
\textit{Lemma}---For the case that $\cos \delta _{1k2l}=+1$, the necessary
and sufficient condition in order for the probability (3a) to vanish is 
\begin{equation}
\tan \beta _{1k}\tan \beta _{2l}\,=\,-c_{1}/c_{2}\, .  \tag{9a}
\end{equation}
Analogously, for the case that $\cos \delta _{1k2l}=+1$, the vanishing of
the probability in Eq.\ (3b) is equivalent to requiring that 
\begin{equation}
\tan \beta _{1k}\tan \beta _{2l}\,=\,-c_{2}/c_{1}\, .  \tag{9b}
\setcounter{equation}{9}
\end{equation}

\vspace{-.3cm}
\noindent
\rule{\textwidth}{.15mm}
\smallskip

For the case of maximal entanglement we have $\left| c_{1}\right| =\left|
c_{2}\right| $, and then, supposing for example that the coefficients $c_{1}$
and $c_{2}$ are of the same sign, both of conditions (9a) and (9b) reduce to
the single one,\footnote{%
Naturally the coincidence of both conditions (9a) and (9b) in the case of
maximal entanglement stems from the fact that, for this case, $P_{\eta
}(D_{1k}=+1,D_{2l}=+1)=P_{\eta }(D_{1k}=-1,D_{2l}=-1).$}
\begin{equation}
\tan \beta _{1k}\tan \beta _{2l}\,=\,-1\, .
\end{equation}
Thus, taking into account that $\delta _{1121}=0$, $\delta _{1122}=0$, and $%
\delta _{1221}=0$, we can apply the previous lemma to conclude that, if
conditions (8a), (8b), and (8c) are to be satisfied for the maximally
entangled state, we must have the relations 
\begin{align}
\tan \beta _{11}\tan \beta _{21}&\,=\,-1\, ,  \tag{11a}  \\
\tan \beta _{11}\tan \beta _{22}&\,=\,-1\, ,  \tag{11b}  \\
\tan \beta _{12}\tan \beta _{21}&\,=\,-1\, .  \tag{11c}  
\end{align}
An immediate but crucial consequence of relations (11a)-(11c) is 
\begin{equation}
\tan \beta _{12}\tan \beta _{22}\,=\,-1\, .   \tag{11d}
\setcounter{equation}{11}
\end{equation}
(This is obtained by simply multiplying Eqs.\ (11b) and (11c), and then
substituting Eq.\ (11a) into the left-hand side of the product.) Relation
(11d), together with the above deduced constraint $\delta _{1222}=0$, allows
one to finally conclude that the fulfillment of all three conditions
(8a)-(8c) by the maximally entangled state necessarily implies that both
probabilities $P_{\eta }(D_{12}=+1,D_{22}=+1)$ and $P_{\eta
}(D_{12}=-1,D_{22}=-1)$ are equal to zero, thus contradicting the condition
in Eq.\ (8d). Clearly, this in turn means that the correlation function $%
E_{\eta }(D_{12},D_{22})$ has to take the extremal value $-1$. Naturally,
the result $E_{\eta }(D_{12},D_{22})=-1$ also follows directly by applying
Eqs.\ (6) and (11d). Indeed, for the case considered we have $2c_{1}c_{2}\cos
\delta _{1222}=+1$, and then, by Eq.\ (6), $E_{\eta }(D_{12},D_{22})=\cos
2(\beta _{12}-\beta _{22})$. Now, from Eq.\ (11d), the parameters $\beta
_{12} $ and $\beta _{22}$ ought to satisfy the relation $\beta _{12}-\beta
_{22}=m_{1222}\pi /2$, with $m_{1222}$ being an odd integer $\pm 1,\pm
3,\ldots $ . Hence the result $E_{\eta }(D_{12},D_{22})=-1$, as claimed.

We will note, incidentally, that the fulfillment of condition (8d) for the
maximally entangled state precludes the simultaneous fulfillment of all
three conditions (8a)-(8c). Indeed, in order to have $E_{\eta
}(D_{12},D_{22})\neq -1$ for the maximally entangled state it is necessary
that (supposing $\delta _{1222}=0$, and sgn~$c_{1}=$ sgn $c_{2}$), 
\begin{equation}
\tan \beta _{12}\tan \beta _{22}\,\neq\, -1\, ,
\end{equation}
and thus at least one of the conditions in Eqs.\ (11a)-(11c) cannot be
satisfied. As an example of this, consider the case where both conditions
(11a) and (11b) are satisfied. Yet, this in turn implies that $\tan \beta
_{22}=\tan \beta _{21}$, and then, from Eq.\ (12) we have that $\tan \beta
_{12}\tan \beta _{21}\neq -1$, in disagreement with Eq.\ (11c).

The results following Eqs.\ (11a)-(11d) can be conveniently summarized as
follows: whenever we have $E_{\eta }(D_{11},D_{21})=E_{\eta
}(D_{11},D_{22})=E_{\eta }(D_{12},D_{21})=-1$ for the maximally entangled
state, then necessarily we must have\linebreak $E_{\eta }(D_{12},D_{22})=-1$
as well. An analogous argument could be established to conclude that,
whenever $E_{\eta }(D_{11},D_{21})=E_{\eta }(D_{11},D_{22})=E_{\eta
}(D_{12},D_{21})=+1$ for the maximally entangled state, then $E_{\eta
}(D_{12},D_{22})=+1$. As we have seen, this very fact makes it impossible
the simultaneous fulfillment of all the Hardy conditions (8a)-(8d) for the
nontrivial case of maximal entanglement (of course, this impossibility holds
trivially for the case of product states).

Let us now consider Bell's inequality in the form of the
Clauser-Horne-Shimony-Holt (CHSH) inequality [2]. This can be written as 
\begin{equation}
\left| E_{\eta }(D_{11},D_{21})+E_{\eta }(D_{11},D_{22})+E_{\eta
}(D_{12},D_{21})-E_{\eta }(D_{12},D_{22})\right| \leq 2\, .
\end{equation}
It is evident that, as it should be expected, the perfect correlations we
have just derived for the maximally entangled state do satisfy the CHSH
inequality. Indeed, for this case we have $\Delta =2$, where the quantity $%
\Delta $ is defined by $\Delta \equiv \left| E_{\eta
}(D_{11},D_{21})+E_{\eta }(D_{11},D_{22})+E_{\eta }(D_{12},D_{21})-E_{\eta
}(D_{12},D_{22})\right| $. The validity of the CHSH inequality for the
simplest case of a $2\times 2$ system described by a pure state is a
necessary and sufficient condition for the existence of a (deterministic)
local hidden-variable model for the observable correlations of all
combinations of measurements $(D_{1k},D_{2l})$ independently performed on
both subsystems [12].\footnote{%
It should be noted, however, that for the case in which the $2\times 2$ system
is described by a mixed state the fulfillment of the CHSH inequality is not in
general a sufficient condition for the existence of a local hidden-variable
model that reproduces the results of more complex (``nonideal'')
measurements. See, for example, the introductory part of the article in Ref.\
13, and references therein.} Therefore, we can say with complete confidence
that all the perfect correlations $E_{\eta }(D_{11},D_{21})=E_{\eta
}(D_{11},D_{22})=E_{\eta }(D_{12},D_{21})=E_{\eta }(D_{12},D_{22})=\pm 1$
can be consistently explained by a local hidden-variable model. Indeed, it
is easy to see that all these perfect correlations are compatible with the
assumption of local realism, so that they can be reproduced by a local
realistic model. So, consider for example the case when all the perfect
correlations above are $-1$, and suppose that a joint measurement of the
observables $D_{11}$ and $D_{21}$ is carried out. Owing to the perfect
correlation $E_{\eta }(D_{11},D_{21})=-1$, the measurement results for such
observables should be, respectively, $+1$ and $-1$ (or else, $-1$ and $+1$).
Suppose now that, instead of $D_{21}$, the observable $D_{22}$ had been
measured on particle 2. According to local realism, the result obtained in a
measurement of $D_{1k}$ on particle 1 cannot depend in any way on which
observable $D_{2l}$ happens to be measured on the other, spatially separated
particle 2, and vice versa. Thus, taking into account the constraint $%
E_{\eta }(D_{11},D_{22})=-1$, we conclude that, had the observable $D_{22}$
been measured, the outcome $-1$ ($+1$) would have been observed whenever the
measurement result for $D_{11}$ is found to be $+1$ ($-1$). Applying a
symmetrical reasoning to the pair of observables $D_{12}$ and $D_{21}$, and
noting that $E_{\eta }(D_{12},D_{21})=-1$, we finally deduce that, had $%
D_{12}$ been measured on particle 1, the outcome $+1$ ($-1$) would have been
observed whenever the measurement result for $D_{21}$ is found to be $-1$ ($%
+1$). In short, by applying local realism, and for the case that the
measurement results for $D_{11}$ and $D_{21}$ happen to be, respectively, $%
+1 $ and $-1$ ($-1$ and $+1$), it is concluded that the measurement results
for $D_{12}$ and $D_{22}$ would have been, respectively, $+1$ and $-1$ ($-1$
and $+1$). But of course this is consistent with the fact that $E_{\eta
}(D_{12},D_{22})=-1$. The compatibility of these perfect correlations with
the assumption of local realism implies that the CHSH inequality should be
satisfied by such perfect correlations, in accordance with our numerical
result that the parameter $\Delta $ has to be equal to 2 for the maximally
entangled state, if this is to satisfy each of the Hardy equations (8a),
(8b), and (8c).\footnote{%
Precisely speaking, what we have done so far is to show that the four 
\textit{given} perfect correlations $E_{\eta }(D_{11},D_{21})=\pm 1$, $%
E_{\eta }(D_{11},D_{22})=\pm 1$, $E_{\eta }(D_{12},D_{21})=\pm 1$, and $%
E_{\eta }(D_{12},D_{22})=\pm 1$ (with each of them taking simultaneously
either the $+$ or $-$ sign) do not contradict each other when local realism
is assumed to hold. On the other hand, Bell's local model for pairs of spin-$%
\frac{1}{2}$ particles in the singlet state [14] (see also Sec.\ II and
Appendix D of Ref.\ 7) provides perhaps the simplest example of a local
hidden-variable model capable of reproducing the perfect correlations $%
E_{\eta }(\mathbf{n}_{1},\mathbf{n}_{2})=\pm 1$ that arise in the special
case that $\mathbf{n}_{1}=\mathbf{n}_{2}$ or $\mathbf{n}_{1}=-\mathbf{n}_{2}$%
, where $\mathbf{n}_{1}$ and $\mathbf{n}_{2}$ are the directions along which
the spin is measured. Therefore, our demonstration supplemented by Bell's
model (or suitable generalizations of it) allows one to fully describe all
the quantum perfect correlations $E_{\eta }(D_{11},D_{21})=E_{\eta
}(D_{11},D_{22})=E_{\eta }(D_{12},D_{21})=E_{\eta }(D_{12},D_{22})=\pm 1$
entirely in terms of a local realistic model.} It is worth noting, however,
that no local hidden-variable model exists that accommodates all the perfect
correlations $E_{\eta }(D_{11},D_{21})=E_{\eta }(D_{11},D_{22})=E_{\eta
}(D_{12},D_{21})=-E_{\eta }(D_{12},D_{22})=\pm 1$, in spite of the fact that
there indeed exists a classical model accounting for each of these perfect
correlations separately. This is because for this case the parameter $\Delta 
$ turns to be greater than 2 (in fact, it attains the maximum possible
theoretical value $\Delta =4$), and then the violation of the CHSH
inequality excludes the existence of a local hidden-variable model that
reproduces simultaneously \textit{all} four correlation functions involved
in it.

It is a well-known fact [15] that the maximum value of $\Delta $ predicted
by quantum mechanics is $2\sqrt{2}$. This means, in particular, that the
maximum possible violation $\Delta =4$ cannot be realized in quantum
mechanics. Clearly, the parameter $\Delta $ takes on the value 4 if, and
only if, $E_{\eta }(D_{11},D_{21})=E_{\eta }(D_{11},D_{22})=E_{\eta
}(D_{12},D_{21})=-E_{\eta }(D_{12},D_{22})=\pm 1$. From the discussion
leading to Eq.\ (7) it follows that, for the general case in which $\beta
_{1k}\neq n_{1k}\pi /2$ and $\beta _{2l}\neq n_{2l}\pi /2$, it is necessary
that the state $\left| \eta \right\rangle $ be maximally entangled, if we
want that the quantum correlation function $E_{\eta }(D_{1k},D_{2l})$
attains the value $+1$ or $-1$. But, as we have shown, the impossibility for
the maximally entangled state to simultaneously satisfy all the Hardy
equations (8a)-(8d) resides in the fact that, whenever we have $E_{\eta
}(D_{11},D_{21})=E_{\eta }(D_{11},D_{22})=E_{\eta }(D_{12},D_{21})=\pm 1$
for such a state, then necessarily $E_{\eta }(D_{12},D_{22})=\pm 1$. The
interesting observation then is that the failure of Hardy's nonlocality
theorem [1] for the maximally entangled case just prevents the quantum
prediction for the parameter $\Delta $ from reaching the maximum theoretical
value $\Delta =4$. Indeed, for this case the quantum prediction for $\Delta $
falls to 2, and then all the relevant quantum correlation functions $E_{\eta
}(D_{11},D_{21})$, $E_{\eta }(D_{11},D_{22})$, $E_{\eta }(D_{12},D_{21})$,
and $E_{\eta }(D_{12},D_{22})$ can be interpreted in terms of a classical
model based on the assumption of local realism.

On the other hand, as already expounded by Krenn and Svozil [16], a maximum
violation of the CHSH inequality by the value 4 would correspond to a
two-particle analog of the GHZ argument for nonlocality [7]. Indeed, as we
have seen, if we have $E_{\eta }(D_{11},D_{21})=E_{\eta
}(D_{11},D_{22})=E_{\eta }(D_{12},D_{21})=\pm 1$, then, according to local
realism, we must have the prediction $E_{\eta }(D_{12},D_{22})=\pm 1$, which
directly contradicts the (hypothetical) result $E_{\eta }(D_{12},D_{22})=\mp
1$. As we are dealing with perfect correlations, this contradiction would
apply to each of the pairs in the ensemble. The failure of Hardy's theorem
for the maximally entangled state can thus also be read as a proof of the
fact that it is not possible to construct a GHZ-type nonlocality argument
for a $2\times 2$ system (unless hypothetical extremely nonclassical
correlations are assumed to hold [16]).

\section{Hardy's nonlocality conditions: implications
for the general case}

Let us now consider the case of less-than-maximally entangled state, that
is, one for which $\left| c_{1}\right| \neq \left| c_{2}\right| $ in Eq.\
(1). For this case the equalities $P_{\eta }(D_{1k}=+1,D_{2l}=+1)=P_{\eta
}(D_{1k}=-1,D_{2l}=-1)$ and $P_{\eta }(D_{1k}=+1,D_{2l}=-1)=P_{\eta
}(D_{1k}=-1,D_{2l}=+1)$ are no longer valid, and then the fulfillment of
conditions (8a), (8b), and (8c) does not imply any perfect correlation
between the measurement results of $D_{11}$ and $D_{21}$, $D_{11}$ and $%
D_{22}$, and $D_{12}$ and $D_{21}$, respectively. According to the lemma
in Sec.\ 3, the fulfillment of the Hardy equations (8a), (8b), and (8c) is
equivalent to requiring, respectively, (as before, we assume that $\delta
_{1121}=\delta _{1122}=\delta _{1221}=0$) 
\begin{align}
\tan \beta _{11}\tan \beta _{21} &\,=\,-c_{2}/c_{1}\, ,  \tag{14a} \\
\tan \beta _{11}\tan \beta _{22} &\,=\,-c_{1}/c_{2}\, ,  \tag{14b} \\
\tan \beta _{12}\tan \beta _{21} &\,=\,-c_{1}/c_{2}\, .  \tag{14c}
\end{align}
From Eqs.\ (14a)-(14c) we get 
\begin{equation}
\tan \beta _{12}\tan \beta _{22}\,=\,-\left( c_{1}/c_{2}\right) ^{3}.
\tag{14d}
\end{equation}
Clearly, whenever $c_{1},c_{2}\neq 0$ and $\left| c_{1}\right| \neq \left|
c_{2}\right| $, we have from Eq.\ (14d) that $\tan \beta _{12}\tan \beta
_{22}\neq -c_{1}/c_{2}$, and, therefore, it is concluded that, for the
nonmaximally entangled case, the fulfillment of conditions (8a)-(8c)
automatically implies the fulfillment of condition (8d). It should be noted
at this point that, for any given $c_{1}$ and $c_{2}$, the Eqs.\ (14a)-(14c)
do not uniquely determine the four parameters $\beta _{ij}$, so that we can
always choose one of them in an unrestricted way [6]. So, in what follows,
we shall take $\beta _{12}$ to be equal to $\beta _{0}$, with $\beta _{0}$
being a variable taking on any \textit{arbitrary} value. Naturally, once the
parameter $\beta _{12}$ is given, the remaining three are forced to
accommodate. Indeed, from Eqs.\ (14a)-(14c), we obtain immediately
\begin{align}
\tan \beta _{11} &\,=\,\left( c_{2}/c_{1}\right) ^{2}\tan \beta _{0}\, ,
\tag{15a} \\
\tan \beta _{21} &\,=\,-(c_{1}/c_{2})\cot \beta _{0}\, ,  \tag{15b} \\
\tan \beta _{22} &\,=\,-(c_{1}/c_{2})^{3}\cot \beta _{0}\, .  \tag{15c}
\setcounter{equation}{15}
\end{align}

We now show explicitly how the fulfillment of all the Hardy conditions
(8a)-(8d) relates to the violation of the CHSH inequality. Since such
conditions cannot all be satisfied within a local and realistic framework,
it is to be expected that the fulfillment of Eqs.\ (8a)-(8d) entails a
violation of the resulting CHSH inequality. That this is indeed the case can be
demonstrated in a rather general way as follows. For this purpose, it is
convenient to express the parameter $\Delta $ in terms of the set of
probabilities $P_{\eta }^{=}(D_{1k},D_{2l})=P_{\eta
}(D_{1k}=+1,D_{2l}=+1)+P_{\eta }(D_{1k}=-1,D_{2l}=-1)$, with $k,l=1,2$.
Provided with such probabilities, the quantity $\Delta $ can be written as 
\begin{equation}
\Delta =2\left| P_{\eta }^{=}(D_{11},D_{21})+P_{\eta
}^{=}(D_{11},D_{22})+P_{\eta }^{=}(D_{12},D_{21})-P_{\eta
}^{=}(D_{12},D_{22})-1\right| .
\end{equation}
Expression (16) is, as it stands, completely general. Now, for the
particular case where conditions (8a)-(8d) are satisfied, we have\footnote{%
Actually, Eq.\ (17) also applies to the case that the probability $P_{\eta
}(D_{12}=+1,D_{22}=+1)$ in Eq.\ (8d) is equal to zero.} 
\begin{eqnarray}
\Delta =2 & \!\!\! \mid & \!\!\!\! P_{\eta }(D_{11}=+1,D_{21}=+1) +
P_{\eta}(D_{11}=-1,D_{22}=-1)   \nonumber \\
& \!\!\!\!\!\! + & \!\!\!\! P_{\eta }(D_{12}=-1,D_{21}=-1)-P_{\eta }
(D_{12}=+1,D_{22}=+1) \nonumber \\
& \!\!\!\!\!\! - &\!\!\!\! P_{\eta }(D_{12}=-1,D_{22}=-1)-1 \! \mid .
\end{eqnarray}
Using Eqs.\ (3a) and (3b) in (17) (with $\cos \delta_{1k2l}=+1$), and taking
into account the constraints in Eqs.\ (14a)-(14d) and (15a)-(15c), we find,
after a bit lengthy but straightforward calculation,
\begin{eqnarray}
\Delta \!\!\!\! &=& \!\!\!\! 2\left| \,\frac{\left( 2c_{1}^{2}-1\right)
^{2}}{1+\frac{\left( 1-c_{1}^{2}\right) ^{2}}{c_{1}^{2}}\tan ^{2}\beta _{0}+%
\frac{c_{1}^{4}}{1-c_{1}^{2}}\cot ^{2}\beta _{0}}\right. +\frac{\left(
2c_{1}^{2}-1\right) ^{2}}{1+\frac{\left( 1-c_{1}^{2}\right) ^{3}}{c_{1}^{4}}%
\tan ^{2}\beta _{0}+\frac{c_{1}^{6}}{\left( 1-c_{1}^{2}\right) ^{2}}\cot
^{2}\beta _{0}}  \nonumber \\
&&\;\;\,\medskip +\,\,\frac{\left( 2c_{1}^{2}-1\right) ^{2}\cos ^{2}\beta
_{0}}{1-c_{1}^{2}+c_{1}^{2}\cot ^{2}\beta _{0}}\,-\,c_{1}^{2}\left( 1-\frac{%
c_{1}^{2}}{1-c_{1}^{2}}\right) ^{2}\frac{\cos ^{2}\beta _{0}}{1+\left( \frac{%
c_{1}^{2}}{1-c_{1}^{2}}\right) ^{3}\cot ^{2}\beta _{0}}  \nonumber \\
&&\;\;\,-\,\,\left( 1-c_{1}^{2}\right) \left. \left( 1-\frac{c_{1}^{4}}{%
\left( 1-c_{1}^{2}\right) ^{2}}\right) ^{2} \frac{\cos ^{2}\beta _{0}}{%
1+\left( \frac{c_{1}^{2}}{1-c_{1}^{2}}\right) ^{3}\cot ^{2}\beta _{0}}%
\,-\,1\,\right| .
\end{eqnarray}
The parameter $\Delta $ given by (18) is represented graphically in Fig.\ 1
as a function of $c_{1}^{2}$ and $\beta _{0}$ for the ranges of variation $0%
\leq c_{1}^{2}\leq 1$ and $0^{\circ }\leq \beta _{0}\leq 90^{\circ }$. From
Fig.\ 1, it can be seen that $\Delta $ is greater than 2 for all values of $%
c_{1}^{2}$ and $\beta _{0}$ except for $c_{1}^{2}=0,1,$ and $0.5$ (that is,
product and maximally entangled states), and $\beta _{0}=n\pi /2$, $n=0,\pm
1,\pm 2,\ldots $ . This latter
\begin{figure}[t]
\captionstyle{centerlast}
\hspace{1.77cm}\includegraphics[width=4.1in]{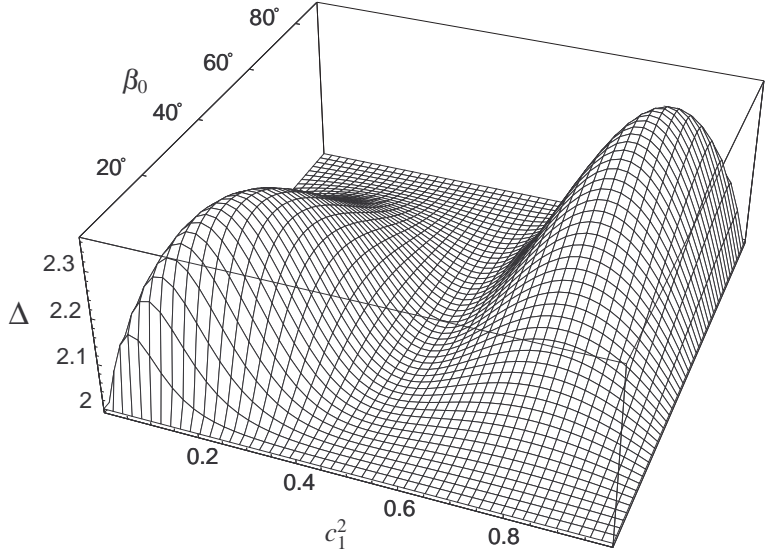}
\setlength{\abovecaptionskip}{0.2cm}
\setlength{\textfloatsep}{9pt plus 2pt minus 3pt}
\setcaptionmargin{1cm}
\renewcommand{\figurename}{Fig.}
\renewcommand{\captionlabeldelim}{.~}
\caption{\footnotesize Quantum prediction for the parameter $\Delta $ in the
case that conditions in Eqs.\ (8a)-(8c) are satisfied. As explained in the text,
the behavior of $\Delta $ is entirely governed by the probability in Eq.\ (8d),
so that $\Delta $ is greater than 2 whenever $P_{\eta }(D_{12}=+1,D_{22}=+1)$
is positive. The surface displayed in the figure fulfills the symmetry
property $\Delta (c_{1}^{2},\beta _{0})=\Delta (1-c_{1}^{2},90^{\circ
}-\beta _{0})$, so that such a surface stays invariant under a rotation of $%
n\pi $ ($n=0,\pm 1,\pm 2,\ldots )$ about a vertical axis passing through the
point $(c_{1}^{2},\beta _{0})=(0.5,45^{\circ })$.}
\vskip1.mm
\end{figure}
exception arises because, if $\beta _{0}=n\pi
/2$, then by Eqs.\ (15a)-(15c) we must have necessarily $\beta
_{ij}=n_{ij}\pi /2$ for each $i,j=1,2$, and for some $n_{ij}=0,\pm 1,\pm
2,\ldots $ (with $n_{12}=n$). In view of Eq.\ (6), this in turn implies that
perfect correlations would take place between the measurement outcomes for
any of the pairs of observables ($D_{1k},D_{2l}$) (see Sec.\ 2). Further, as
may be easily checked, these perfect correlations fulfill $E_{\eta
}(D_{11},D_{21})=E_{\eta }(D_{11},D_{22})=E_{\eta }(D_{12},D_{21})=E_{\eta
}(D_{12},D_{22})=-1$, and thus $\Delta =2$. It is not difficult to show, on
the other hand, that the expression (18) remains invariant under the joint
transformations $c_{1}^{2}\rightarrow c_{2}^{2}=1-c_{1}^{2}$ and $\beta
_{0}\rightarrow \pm \beta _{0}+m\pi /2$, $m=\pm 1,\pm 3,\ldots $ . In fact,
the greatest value of $\Delta $ is attained for both sets of points $%
(c_{1}^{2},\beta _{0})=(0.177\,\,352,\pm 17.5566^{\circ }+n\pi )$ and $%
(c_{1}^{2},\beta _{0})=(0.822\,\,648,\pm 72.4434^{\circ }+n\pi )$,
where $n=0,\pm 1,\pm 2,\ldots $ . This greatest value is $\Delta _{g}=2.360\,
\,679$ or, in closed notation, $2+4\tau ^{-5}$, with $\tau $ being the golden
mean $\frac{1}{2}(1+\sqrt{5})$. The quantity $\Delta _{g}-2$ corresponds to
approximately 43.5\% of the maximum violation $2\sqrt{2}-2$ predicted by
quantum mechanics of the CHSH inequality.

It is worthwhile to mention that the graph for $\Delta $ in Fig.\ 1 has quite
the same shape as the graphical representation of the probability $P_{\eta
}(D_{12}=+1,D_{22}=+1)$ in Eq.\ (8d) (this latter graph can be found in Ref.\
6), the only relevant difference being the respective ranges of variation
of the values taken by such functions, namely, $[2,2+4\tau ^{-5}]$ for $%
\Delta $, and $[0,\tau ^{-5}]$ for $P_{\eta }(D_{12}=+1,D_{22}=+1)$. Indeed,
we have proved with the aid of a computer program that the whole expression
(18) is connected with the probability function $P_{\eta
}(D_{12}=+1,D_{22}=+1)$ (which is given explicitly by the fourth term on the
right-hand side of Eq.\ (18)) through the simple relation 
\begin{equation}
\Delta =2+4P_{\eta }(D_{12}=+1,D_{22}=+1)\, .
\end{equation}
In particular this means that the parameter $\Delta $ in Eq.\ (18) is maximum
whenever $P_{\eta }(D_{12}=+1,D_{22}=+1)$ so is. Likewise, $\Delta $ takes
its minimum value 2 whenever $P_{\eta }(D_{12}=+1,D_{22}=+1)$ vanishes. This
close relationship between $\Delta $ and $P_{\eta }(D_{12}=+1,D_{22}=+1)$
was to be expected since the value of $P_{\eta }(D_{12}=+1,D_{22}=+1)$ can
be regarded as a direct measure of the degree of ``nonlocality'' inherent
in the Hardy equations (8a)-(8d).

It will further be noted, incidentally, that Hardy's argument for
nonlocality can equally be cast in the form of a simple inequality involving
the four probabilities in Eqs.\ (8a)-(8d) [17]:
\begin{eqnarray}
P_{\eta }(D_{12}=+1,D_{22}=+1)\leq \!\!\!\!\!\!\!\! &&P_{\eta
}(D_{11}=-1,D_{21}=-1)  \nonumber \\
+\!\!\!\!\!\!\!\! &&P_{\eta }(D_{11}=+1,D_{22}=+1)  \nonumber \\
+\!\!\!\!\!\!\!\! &&P_{\eta }(D_{12}=+1,D_{21}=+1)\, .
\end{eqnarray}
Quantum mechanics predicts a maximum violation of inequality (20) for the
values $P_{\eta }(D_{11}=-1,D_{21}=-1)=P_{\eta
}(D_{11}=+1,D_{22}=+1)=P_{\eta }(D_{12}=+1,D_{21}=+1)=0$, and $P_{\eta
}(D_{12}=+1,D_{22}=+1)=\tau ^{-5}$. The point to be stressed here is that,
for this same set of values, quantum mechanics predicts a violation of the
CHSH inequality which is four times bigger than that obtained for the
inequality (20). It is therefore concluded that, in order to achieve a more
conclusive, clear-cut experimental verification of Hardy's nonlocality
theorem [1], one could try to measure the observable probabilities in Eq.\ (17),
once the conditions $P_{\eta }(D_{11}=-1,D_{21}=-1)=P_{\eta}(D_{11}=+1,D_{22}
=+1)=P_{\eta }(D_{12}=+1,D_{21}=+1)=0$, and $P_{\eta }(D_{12}=+1,D_{22}=+1)=
\tau^{-5}$ have been established.\footnote{%
Experimentally, it is not possible to achieve in any case a true ``zero''
value for the various probabilities $P_{\eta }(D_{1k}=m,D_{2l}=n)$, these
values remaining necessarily finite. As an illustration of this, we may quote
the experimental results corresponding to the probabilities in Eqs.\ (8a)-(8d)
obtained in the first actual test of Hardy's theorem [18]. This is a
two-photon coincidence experiment, and the reported results are $P(\theta
_{10},\theta_{20})=0.0070\pm 0.0005$, $P(\theta_1,\bar{\theta}_{20})=0.0034
\pm 0.0004$, $P(\bar{\theta}_{10},\theta_2)=0.0040\pm 0.0004$, and $P(\theta_1,
\theta_2)=0.099\pm 0.002$, which were obtained for the following polarizer
angles, $\theta_1=74.3^{\circ}$, $\theta_2=15.7^{\circ}$, $\theta_{10}=-56.8^
{\circ}$, $\bar{\theta}_{10}=33.2^{\circ}$, $\theta_{20}=-33.2^{\circ}$, and
$\bar{\theta}_{20}=56.8^{\circ}$. As can be seen, the first three quoted
probabilities are close to zero, while the fourth one is significantly
different from zero. Subsequent experimental work on Hardy's theorem is
reported in Ref.\ 19.}

Bell inequalities should be satisfied by any realistic theory fulfilling a
very broad and general locality condition according to which the real
factual situation of a system must be independent of anything that may be
done with some other system which is spatially separated from, and not
interacting with, the former [8,\,20]. When applied to our particular
situation, this requirement essentially means that the effect of the choice
of the observable $D_{i1}$ or $D_{i2}$ to be measured on particle $i$, $i=1,2$,
cannot influence the result obtained with another remote measuring device
acting on the other particle. For this class of theories the ensemble
(measurable) probability of jointly obtaining the result $D_{1k}=\pm 1$ for
particle 1 and the result $D_{2l}=\pm 1$ for particle 2, has the functional
form [2,\,21,\,22] 
\begin{equation}
P_{\text{HV}}(D_{1k}=\pm 1,D_{2l}=\pm 1)=\int_{\Lambda }d\lambda \, \rho
(\lambda )P\left( D_{1k}=\pm 1\right| \lambda )P\left( D_{2l}=\pm 1\right|
\lambda )\, .
\end{equation}
In Eq.\ (21), $\lambda $ is a set of variables (with domain of variation $%
\Lambda $) representing the complete physical state of each individual pair of
particles 1 and 2 emerging from the source, $\rho (\lambda )$ is the (normalized)
hidden-variable distribution function for the initial joint state of the
particles, and $P(D_{ij}=\pm 1\left| \lambda \right) $ is the probability that
an individual particle $i$ in the state $\lambda $ gives the result $\pm 1$ for a
measurement of $D_{ij}$. Note that both $\rho (\lambda )$ and $\Lambda $ are
independent of the actual setting $j=1$ or $2$ corresponding, respectively, to a
measurement of $D_{i1}$ or $D_{i2}$ on particle $i$. As is well known,
classical probabilities of the form (21) lead to validity of the
inequality $\Delta \leq 2$ (and, generally speaking, to validity of any
other Bell-type inequality). This is usually proved by invoking certain
algebraic theorems (see, for instance, Ref.\ 22 for a derivation of Bell's
inequality in the context of actual optical tests of local hidden-variable
theories). Since the fulfillment of all the Hardy conditions (8a)-(8d)
implies the quantum mechanical violation of the inequality $\Delta \leq 2$
(see Fig.\ 1), it is concluded that no set of probabilities of the form (21)
exists which generally reproduces the quantum prediction (18). The most
remarkable exception to this statement is the case where $\left|
c_{1}\right| =\left| c_{2}\right| $.\footnote{%
Recently, Barnett and Chefles (see Ref.\ 23) have shown how Hardy's
original theorem can be extended to reveal the nonlocality of all pure
entangled states without inequalities. This is accomplished by considering
generalized measurements (that is, measurements beyond the standard von
Neumann type considered here) which perform unambiguous discrimination
between nonorthogonal states.} For this case quantum mechanics predicts $%
\Delta =2$, and then, as was discussed in Sec.\ 3, a rather trivial classical
model of the type considered can be constructed which accounts for each of
the quantum perfect correlations $E_{\eta }(D_{11},D_{21})=E_{\eta
}(D_{11},D_{22})=E_{\eta }(D_{12},D_{21})=E_{\eta }(D_{12},D_{22})=\pm 1$.%
\footnote{%
It is to be noticed that Bell's illustrative model in Ref.\ 14 is a
deterministic one in the sense that the hidden variable $\lambda $ (which, in
Bell's concrete model, is a unit vector in three-dimensional space) uniquely
determines the outcome for any spin measurement on either particle. Eq.\ (21)
above, on the other hand, defines a less restrictive (and, therefore, more
general) type of local hidden-variable theory which is characterized by the
fact that now the set of hidden variables $\lambda $ describing the joint
state of the particles only determines the probability $P(D_{ij}=\pm 1\left|
\lambda \right) $ of obtaining a result $\pm 1$ when the observable $D_{ij}$
is measured on particle $i$, $i=1,2$.}

\section{Concluding remarks}

A final comment is in order about the fact that maximally entangled states
yield the maximum quantum mechanical violation of Bell's inequality, while
they are unable to exhibit Hardy-type nonlocality. The explanation for this
seeming contradiction simply relies on the fact that the rather stringent
constraints (11a)-(11d) implied by the Hardy equations (8a)-(8c) in the case
of a maximally entangled state, are not at all present in the derivation of
Bell's inequality. Indeed, in the case of Bell's theorem, all the parameters 
$\beta _{1k}$, $\beta _{2l}$, $\delta _{1k}$, and $\delta _{2l}$ ($k,l=1,2$)
are treated as independent variables which can assume any arbitrary value
regardless of the quantum state $\left| \eta \right\rangle $ at issue, so
that the Bell inequality will in fact be maximally violated for a suitable
choice of $\beta _{ij}$ and $\delta _{ij}$ (provided $\left| c_{1}\right|
=\left| c_{2}\right| $). So, consider the case in which $2c_{1}c_{2}\cos
\delta _{1k2l}=+1$ for each $k$ and $l$. For this case the quantum
prediction for $\Delta $ is given by 
\begin{equation}
\Delta =\left| \cos 2(\beta _{11}-\beta _{21})+\cos 2(\beta _{11}-\beta
_{22})+\cos 2(\beta _{12}-\beta _{21})-\cos 2(\beta _{12}-\beta
_{22})\right| ,
\end{equation}
which attains the value $\Delta =2\sqrt{2}$ whenever $\beta _{11}-\beta
_{21}=-\pi /8$, $\beta _{11}-\beta _{22}=\pi /8$, $\beta _{12}-\beta
_{21}=\pi /8$, and $\beta _{12}-\beta _{22}=3\pi /8$. Only if the parameters 
$\beta _{ij}$ are constrained to obey the relations (11a)-(11d), as demanded
by the Hardy equations (8a)-(8c) in the case of maximal entanglement, we
have that $\beta _{1k}-\beta _{2l}=m_{1k2l}\pi /2$ for each $k$ and $l$, and
for some odd integer $m_{1k2l}$ (for instance, $\beta _{11}-\beta _{21}=-\pi
/2$, $\beta _{11}-\beta _{22}=\pi /2$, $\beta _{12}-\beta _{21}=\pi /2$, and 
$\beta _{12}-\beta _{22}=3\pi /2$), and then $\Delta =2$.\footnote{%
Of course a similar conclusion applies to the case that sgn $c_{1}\neq $ sgn 
$c_{2}$, so that $2c_{1}c_{2}\cos \delta _{1k2l}=-1$. For this case quantum
mechanics predicts, $\Delta =\left| \,\cos 2(\beta _{11}+\beta _{21})\right.
+\cos 2(\beta _{11}+\beta _{22})+\cos 2(\beta _{12}+\beta _{21})-\left. \cos
2(\beta _{12}+\beta _{22})\, \right| $. Now the fulfillment of Eqs.\ (8a)-(8c)
for the maximally entangled state requires that $\tan \beta _{1k}\tan \beta
_{2l}=1$ for each $k$ and $l$. This in turn implies that $\beta _{1k}+\beta
_{2l}=m_{1k2l}\pi /2$ for some odd integer $m_{1k2l}$, and thus $\Delta =2$.}
Consider now the particular case in which $\left| c_{1}\right| =\left|
c_{2}\right| =2^{-1/2}$, $\beta _{11}=\pi /4$, $\beta _{12}=-3\pi /4$, $%
\beta _{21}=3\pi /4$, and $\beta _{22}=-\pi /4$ (with the parameters $\delta
_{1k}$ and $\delta _{2l}$ taking on any arbitrary value). For this case the
quantum prediction for $\Delta $ becomes 
\begin{equation}
\Delta =\left| \cos (\delta _{11}-\delta _{21})+\cos (\delta _{11}-\delta
_{22})+\cos (\delta _{12}-\delta _{21})-\cos (\delta _{12}-\delta
_{22})\right| ,
\end{equation}
which attains the value $2\sqrt{2}$ for $\delta _{11}-\delta _{21}=-\pi /4$, 
$\delta _{11}-\delta _{22}=\pi /4$, $\delta _{12}-\delta _{21}=\pi /4$, and $%
\delta _{12}-\delta _{22}=3\pi /4$. However, the fulfillment of the Hardy
equations (8a), (8b), and (8c) requires, respectively, that $\delta
_{1121}=n_{1121}\pi $, $\delta _{1122}=n_{1122}\pi $, and $\delta
_{1221}=n_{1221}\pi $. In particular, the choice $\delta _{1121}=\delta
_{1122}=\delta _{1221}=0$ entails that $\delta _{1222}=0$, and therefore,
for such values, the quantity $\Delta $ takes again the value 2.

To summarize, we have shown that whenever the Hardy equations (8a)-(8c) are
fulfilled for the maximally entangled state then perfect correlations develop
between the measurement outcomes $D_{1k}=m$ and $D_{2l}=n$ obtained in any one
of the four possible combinations of joint measurements ($D_{1k},D_{2l}$) one
might actually perform on both particles. As a result, for such observables
$D_{ij}$, the quantity $\Delta $ turns out to be equal to 2, and hence no
violations of local realism will arise in those circumstances. This is in
contrast with the situation entailed by Bell's theorem (with inequalities)
where no constraints such as Eqs.\ (8a)-(8c) need be fulfilled, and then all
the relevant parameters can be varied freely. On the other hand, for the
nonmaximally entangled case, we have generally shown that the fulfillment of
conditions (8a)-(8c)\footnote{%
Remember that the fulfillment of conditions (8a)-(8c) for the nonmaximally
entangled state automatically entails the fulfillment of the remaining
condition in Eq.\ (8d), provided $\beta_{0} \neq n\pi/2$.} 
necessarily makes the parameter $\Delta $ greater than 2, the greatest value
of $\Delta $ predicted by quantum mechanics being as large as $\Delta _{g}=
2+4\tau ^{-5}$. As was emphasized in Sec.\ 4, this result could have some
relevance from an experimental point of view, since it indicates that
experiments based on the inequality $\Delta \leq 2$ (with $\Delta $
given by Eq.\ (17)) would be more efficient in order to exhibit Hardy's
nonlocality than those based on inequality (20).\bigskip

\textbf{Acknowledgments} --- The author wishes to thank Agust\'{\i}n del Pino
for his interest and many useful discussions on the foundations of quantum
mechanics. He would also like to thank an anonymous referee for his valuable
suggestions which led to an improvement of an earlier version of this
paper.
\pagebreak

\begin{center}
\textbf{APPENDIX}
\end{center}

The demonstration of the lemma is as follows. Here we give only the proof
that the vanishing of the probability function (3a) for the case that $\cos
\delta _{1k2l}=+1$, is equivalent to the fulfillment of relation (9a), the
proof concerning the equivalence of relation (9b) and the vanishing of Eq.\
(3b) being quite similar. We first show necessity, namely, that the
vanishing of the probability in Eq.\ (3a) implies relation (9a), provided
that $\cos \delta _{1k2l}=+1$. So, equating expression (3a) to zero, and
putting $\cos \delta _{1k2l}=+1$, we get 
\begin{gather}
c_{1}^{2}\cos ^{2}\beta _{1k}\cos ^{2}\beta _{2l}+c_{2}^{2}\sin ^{2}\beta
_{1k}\sin ^{2}\beta _{2l} \nonumber    \\
\,\,=-2c_{1}c_{2}\cos \beta _{1k}\cos \beta _{2l}\sin \beta _{1k}\sin \beta
_{2l}\, ,  \tag{A1}
\end{gather}
or, equivalently, 
\begin{equation}
(c_{1}/c_{2})^{2}\left[ \tan \beta _{1k}\tan \beta _{2l}\right] ^{-1}+\tan
\beta _{1k}\tan \beta _{2l}=-2(c_{1}/c_{2})\, .  \tag{A2}
\end{equation}
Now, making the identifications $\tan \beta _{1k}\tan \beta _{2l}\equiv x$
and $c_{1}/c_{2}\equiv a$, Eq.\ (A2) can be rewritten in the form 
\begin{equation}
x^{2}+2ax+a^{2}=0\, ,  \tag{A3}
\end{equation}
or, 
\begin{equation}
\left( x+a\right) ^{2}=0\, .  \tag{A4}
\end{equation}
Obviously, Eq.\ (A4) is satisfied only for $x=-a$, and, therefore, it is
concluded that the vanishing of the probability (3a) (with $\cos \delta
_{1k2l}=+1$) necessarily entails that $\tan \beta _{1k}\tan \beta
_{2l}=-c_{1}/c_{2}$.

The proof of sufficiency, namely, that the fulfillment of relation (9a)
implies the vanishing of the probability (3a) when $\cos \delta _{1k2l}=+1$,
is quite immediate, and it will not be detailed here.
\pagebreak

\end{document}